\documentclass[aps,prb,preprint,english,a4paper,floatfix,showpacs,10pt]{revtex4-1}
\usepackage[T1]{fontenc}
\usepackage{ae,aecompl}
\usepackage[english]{babel}
\usepackage{mathtools}
\usepackage{url}
\usepackage{psfrag,color,pstricks,pst-grad}
\usepackage[dvips]{graphicx}
\usepackage{amsmath,amsfonts,amssymb,amsbsy}  
\usepackage[normalem]{ulem}
\usepackage{hyperref}

\newcommand{\bb}{\mathbf}
\newcommand{\ca}{\mathcal}

\newcommand{\fbbg}{\left[f_B(\omega,\Tenv)+\frac{1}{2} \right]}
\newcommand{\fbi}{\left[f_B(\omega,T_{i})+\frac{1}{2} \right]}
\newcommand{\fbj}{\left[f_B(\omega,T_{j})+\frac{1}{2} \right]}

\newcommand{\gem}{\mathbb{G}}

\newcommand{\bu}{\bb{u}}
\newcommand{\bE}{\bb{E}}
\newcommand{\br}{\bb{r}}
\newcommand{\bp}{\bb{p}}
\newcommand{\bG}{\bb{G}}
\newcommand{\pom}{(\omega)}
\newcommand{\epsenv}{\varepsilon_{\textrm{env}}}
\newcommand{\Tenv}{T_{\textrm{env}}}
\newcommand{\Eenv}{\bb{E}_{\textrm{env}}}
\newcommand{\Eenvhat}{\hat{\bb{E}}_{\textrm{env}}}
\newcommand{\gemfull}{{\gem}^{\textrm{CM}}}
\newcommand{\tildegemfull}{{\gem}^{\textrm{full}}}
\newcommand{\kw}{k_0}
\newcommand{\unitdyadic}{\bb{I}_{3 \times 3}}
\begin{document}

\title{Quantum Langevin equation approach to electromagnetic energy transfer between dielectric bodies in an inhomogeneous environment}
\author{K. S\"a\"askilahti}
\email{kimmo.saaskilahti@aalto.fi}
\author{J. Oksanen}
\author{J. Tulkki}
\affiliation{Department of Biomedical Engineering and Computational Science, Aalto University, FI-00076 Aalto, Finland}
\date{\today}

\begin{abstract}
Near-field and resonance effects have a strong influence on the nanoscale electromagnetic energy transfer, and detailed understanding of these effects is required for the design of new, optimized nano-optical devices. We provide a comprehensive microscopic view of electromagnetic energy transfer phenomena by introducing quantum Langevin heat baths as local noise sources in the equations of motion for the thermally fluctuating electric dipoles forming dielectric bodies. The theory is, in a sense, the microscopic generalization of the well-known fluctuational electrodynamics theory and thereby provides an alternative and conceptually simple way to calculate the local emission and absorption rates from the local Langevin bath currents. We apply the model to study energy transfer between silicon carbide nanoparticles located in a microcavity formed of two mirrors and next to a surface supporting propagating surface modes. The results show that the heat current between the dipoles placed in a cavity oscillates as a function of their position and distance and can be enhanced by several orders of magnitude as compared to the free space heat current with a similar interparticle distance. The predicted enhancement can be viewed as a many-body generalization of the well-known cavity Purcell effect. Similar effects are also observed in the interparticle heat transfer between dipoles located next to a surface of a polar material supporting surface phonon polaritons. 
\end{abstract}
 \maketitle

\section{Introduction}
\label{sec:intro}

Electromagnetic energy transfer between dielectric bodies at different temperatures is commonly described using the fluctuational electrodynamics (FED) approach \cite{joulain05,volokitin07} developed by Rytov \cite{rytov58,rytov} and first applied to condensed matter physics by Lifshitz \cite{lifshitz55,lifshitz56}. According to FED, thermal motion of charged particles in a body creates random currents, which induce electromagnetic fields. Outside the body, the field is then either radiated to free space or absorbed in the near or far-field regime by another body. Proximity effects involving the evanescent waves in the near-field were first observed by Hargreaves \cite{hargreaves69}, and FED was consequently applied to theoretically predict strong near-field enhancement of heat transfer in various geometries \cite{polder71,loomis94,pendry99,mulet01,volokitin01}. The predictions have been explored in more detail also experimentally \cite{muller-hirsch99,kittel05,hu08,rousseau09,shen09,ottens11}. Near-field effects are expected to have numerous applications in, e.g., thermal microscopy \cite{majumdar99,muller-hirsch99,kittel05,kittel08}, infrared thermophotovoltaics \cite{dimatteo01,narayanaswamy03,laroche06} and narrow-band infrared antennas \cite{carminati99,shchegrov00,greffet02}.


As a statistical model, FED is closely related to Langevin dynamics commonly applied to describe the random thermal motion of non-charged bodies \cite{zwanzig,chaikin}. In Langevin dynamics the particle is assumed to be coupled to a bath of harmonic oscillators, whose effect on the particle can effectively be described by a random force and deterministic friction \cite{ford65,caldeira81,ford88,dhar03,dhar06}. Huttner and Barnett \cite{huttner92} essentially applied Langevin dynamics to study the quantization of the electromagnetic field in absorbing dielectrics by coupling the polarization field to a bath of harmonic oscillators. The explicit inclusion of the microscopic degrees of freedom responsible for absorption solves the problem of temporally decaying field commutators arising if one blindly applies the standard electromagnetic field quantization methods to dielectrics. Relation of the Langevin dynamics to FED was recently highlighted by Rosa, Dalvit and Milonni \cite{rosa10,rosa11}, who used Langevin dynamics to derive the fluctuation-dissipation theorem (FDT) \cite{rytov,lifshitz55,lifshitz56} for the fluctuating polarization field from the microscopic motion of the oscillating dipoles. 

The goal of this paper is to show that the microscopic dipole oscillator model combined with quantum Langevin dynamics can be used to transparently and microscopically treat also the problem of electromagnetic heat transfer between dielectric bodies held at different temperatures. The primary motivation for studying heat transfer using the quantum Langevin equation instead of FED is that we can derive heat transfer rates in full analogy with phononic \cite{dhar03,dhar06,saaskilahti13} and electronic \cite{dhar03} heat transfer and, by following the mathematical manipulations presented in Ref. \cite{saaskilahti13}, we are able to arrive at an identical Landauer-B\"uttiker-like formula for the energy transmission function. The theory presented here enables, therefore, a unification of phononic, electronic and photonic heat transfer under the common Langevin theory.  

When written in terms of particle polarizabilities and the electromagnetic Green's function, the transmission function reduces to the form derived earlier from FED \cite{benabdallah11,messina13}. In contrast to these works, we (i) consistently include the electromagnetic self-interaction produced by the local electromagnetic Green's dyadic by following the discrete dipole approximation  \cite{purcell73,yaghjian80,yurkin07,novotny}, (ii) include the inhomogeneous environment enabling, e.g., accounting for cavity resonance effects, and (iii) present an alternative and conceptually simple way to derive electromagnetic energy transfer rates starting from the microscopic equations of motion. For presentational simplicity, we initially assume the particles to be small enough for the dipole approximation to hold. However, overcoming this assumption by following the well-established discrete dipole approximation mentioned above is also discussed. 

As an application of the formalism, we study the enhancement of heat transfer rates between SiC particles placed in a microcavity and close to a polar surface supporting surface phonon polaritons (SPPs). In a microcavity, the heat transfer rate between particles is shown to oscillate as a function of their distance and the cavity enhancement can be several orders of magnitude as compared to the free space heat current with a similar interparticle distance. Enhanced heat current is predicted also for particles close to a SiC surface, where the SPPs transport electromagnetic energy between the particles.

The paper is organized as follows. In Sec. \ref{sec:theory}, we present the formulation to calculate thermal energy transfer between dielectric particles and the environment. We (i) represent the polarization fields inside the particles by their total dipole moments, (ii) solve the quantum Langevin equations of motion for coupled dipole moment dynamics in terms of the dipole displacement Green's function, (iii) calculate the thermal averages of heat currents using the fluctuation-dissipation theorem for the bath noises and the thermal field of the environment, and (iv) express the heat currents in terms of particle polarizabilities and the electromagnetic Green's dyadic. In Secs. \ref{sec:results_cavity} and \ref{sec:results_surface}, we investigate heat transfer between SiC nanoparticles in a microcavity and above a SiC surface, respectively. 

\section{Theory}
\label{sec:theory}

In this section we formulate the many-particle electromagnetic heat transfer problem in terms of the dipole approximation \cite{novotny} and Langevin dynamics (LD). Our final results for the energy transfer rates are essentially equivalent to those obtained using fluctuational electrodynamics (FED) \cite{benabdallah11,messina13}. Therefore readers only interested in the numerical results regarding the energy transfer rates in inhomogeneous environments are not required to go through the rigorous derivations of this section to understand the results. However, our derivation based on LD provides an alternative view to the FED approach, offering additional insight to the physics involved. In contrast to the FED approach, where the electric fields and the induced dipole moments arising from the fluctuating dipole moments are solved to calculate the energy dissipated by the induced currents, LD is based on describing fluctuations by stochastic forces in the microscopic dipole equations of motion \cite{rosa10}. With the help of Poynting's theorem, we show that the locally absorbed power can be calculated in LD simply from the steady-state energy current to the local Langevin heat bath. Other advantages of the Langevin approach compared to FED are discussed below at the end of Sec. \ref{sec:transmission_polarizability}.



\begin{figure}
  \includegraphics[width=.49\columnwidth]{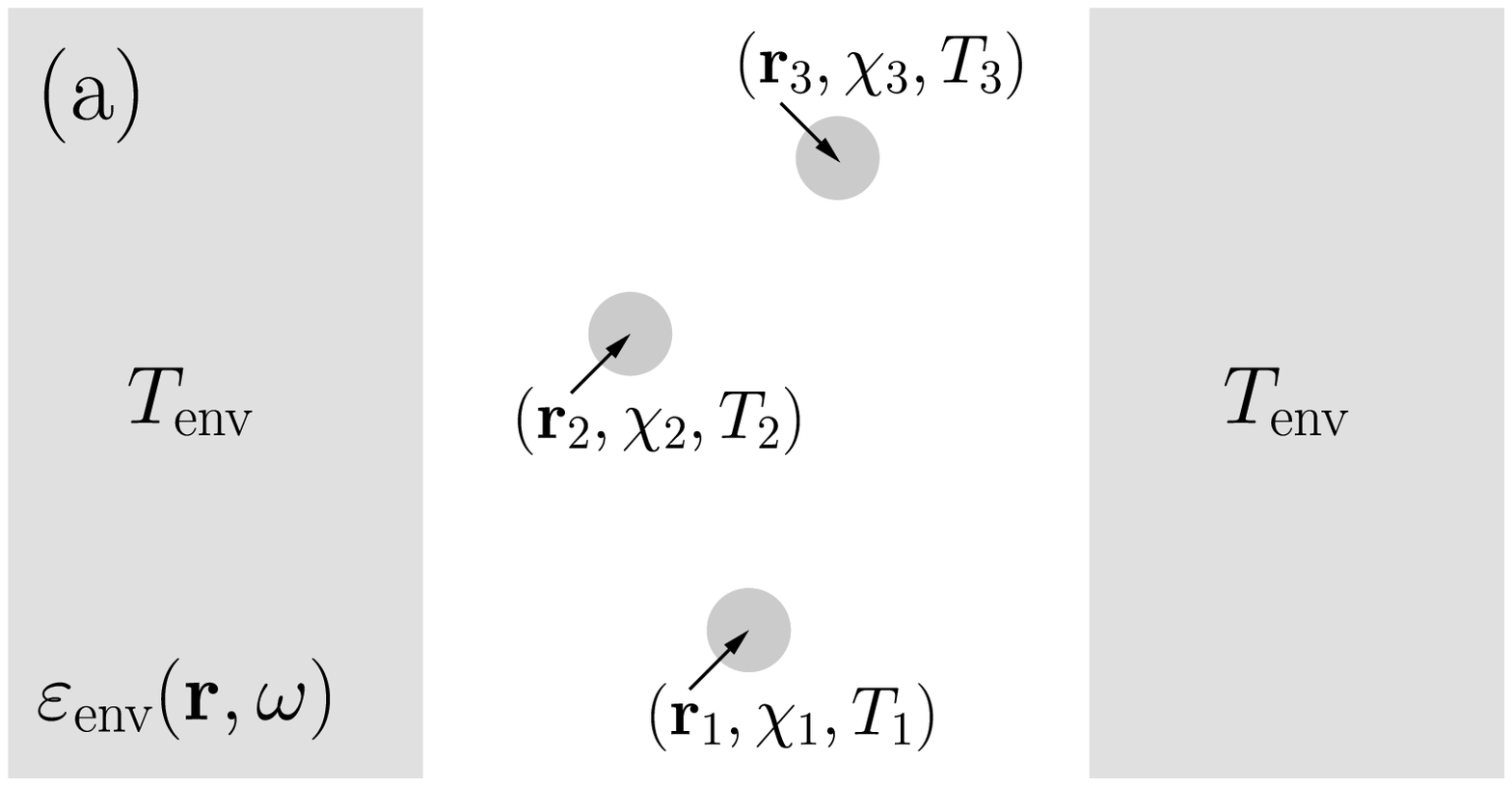}
   \includegraphics[width=.49\columnwidth]{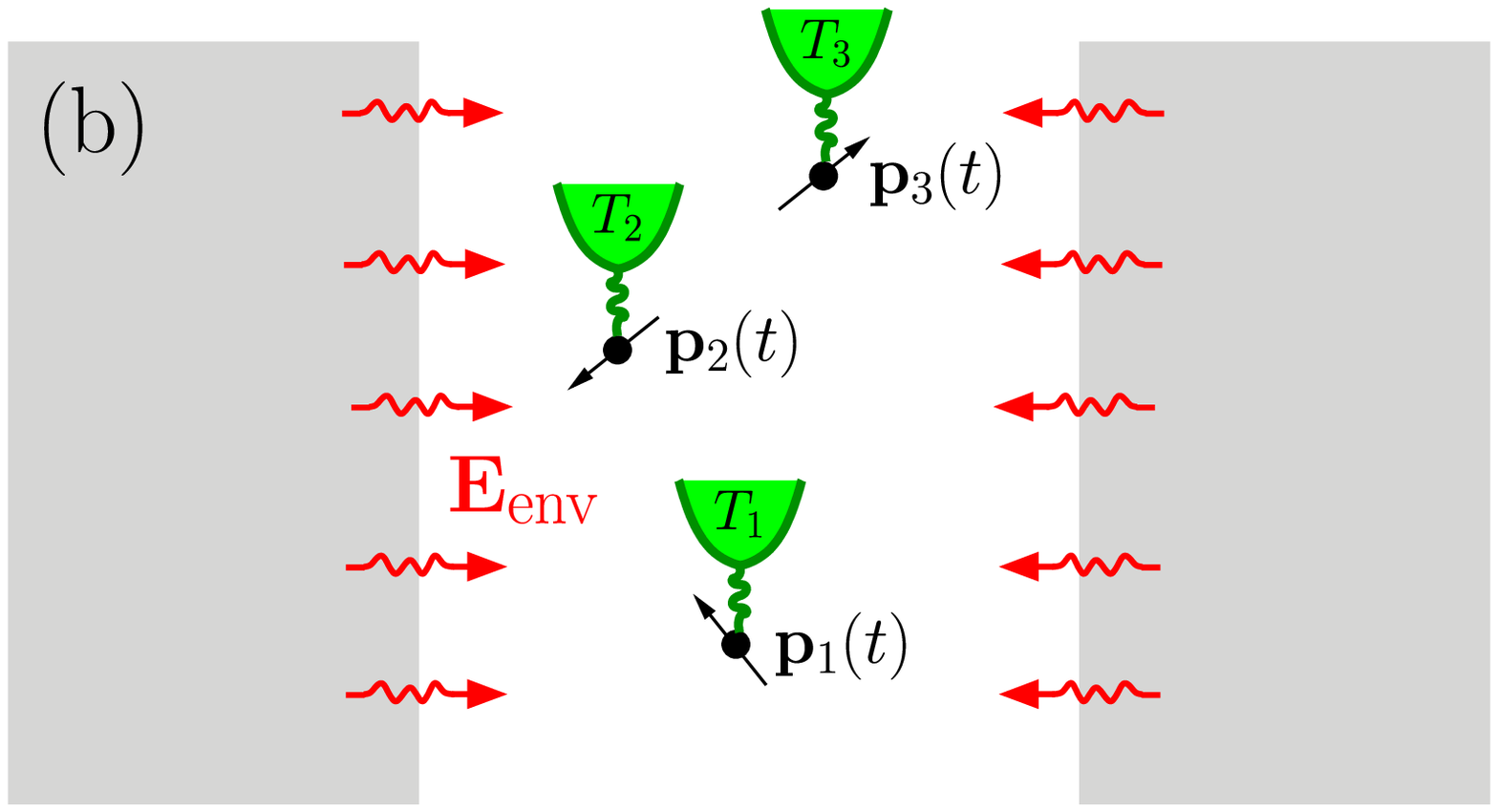}
 \caption{(Color online) (a) A schematic example of the studied system. A collection of $N$ small dielectric particles with positions $\mathbf{r}_i$, electric susceptibilities $\chi_i(\omega)$ and temperatures $T_i$ is located in an inhomogeneous environment, consisting in the shown case of two dielectric (or metallic) bodies occupying the left and right half-spaces. The overall relative permittivity is given by $\varepsilon(\br,\omega)=\epsenv(\br,\omega)$ outside the particles and $\varepsilon(\br_i,\omega)=1+\chi_i(\omega)$ at each particle coordinate $\br_i$. The part described by the environment dielectric constant, consisting in this example of the two cavity walls, is assumed to act as a source of thermal radiation at temperature $\Tenv$. The polarization field inside each particle $i$ is treated as an oscillating point dipole moment $\bb{p}_i$ coupled to a local Langevin bath at temperature $T_i$ as shown in (b). The total local field $\bE(\br_i,t)$ driving each dipole moment $\bb{p}_i$ is the sum of the stochastic background field $\bE_{\textrm{env}}(\br_i,t)$ and the fields $\bE_{ij}(t)$ created by each dipole $j$.}
\label{fig:sud1}
\end{figure}

In our system setup, the studied dielectric particles with electric susceptibilities $\chi_i$ are located in an environment defined through its relative permittivity $\epsenv(\br,\omega)$, so that the overall relative permittivity is given $\varepsilon(\br,\omega)=\epsenv(\br,\omega)$ outside the particles and $\varepsilon(\br_i,\omega)=1+\chi_i(\omega)$ at particle coordinate $\br_i$ as illustrated in Fig. \ref{fig:sud1}(a). Consequently the environment, consisting in the example shown in Fig. \ref{fig:sud1}(a) of the two cavity walls, is described as a single object whose temperature is constant but the permittivity can be inhomogeneous. If the particles are located in pure vacuum environment so that $\epsenv(\br,\omega)=1$ $\forall \bb{r}\in \mathbb{R}^3$, the environment only acts as a source of black-body thermal radiation. If the environment is inhomogeneous, it not only generates background thermal radiation but also scatters the radiation emitted by the dipoles. For simplicity, we assume that all particles are non-magnetic such that the relative magnetic permeability equals unity everywhere.

Following the dipole approximation, the internal polarization field of each particle with susceptibility $\chi_i(\omega)$ is modeled as a dipole located at the central coordinate $\br_i$ of the particle $i$ as illustrated in Fig. \ref{fig:sud1}(b). The microscopic dipole moments, which represent the fluctuating electric polarization inside each particle, are then coupled to (i) local heat baths describing thermal fluctuations and dissipation, (ii) to the electromagnetic field arising from other dipoles, and (iii) to the thermal field originating from the environment.

The dipole formulation results in equations of motion for the dipole displacements, which are solved by using their Fourier transform in Sec. \ref{sec:equationofmotion}. The electromagnetic Green's dyadic coupling the dipoles is defined in Sec. \ref{sec:greens_dyadics}. Combined with the microscopic expressions for local absorption and emission derived from the Poynting theorem in Sec. \ref{sec:poynting}, the solutions of the equations of motion and the correlation functions of the Langevin noise and the environment thermal field described in Sec. \ref{sec:noise} are then used in Sec. \ref{sec:lb_formula} to derive concise expressions for the electromagnetic energy transfer between dipoles as well as the environment field. Finally, in Secs. \ref{sec:polarizabilities} and \ref{sec:transmission_polarizability} we relate the local oscillator parameters to the particle polarizabilities and express the electromagnetic energy transfer rates in terms of these quantities and the full electromagnetic Green's dyadic. 

The dipole approximation is strictly valid if the particles are much smaller than the dominant wavelength and the electric field amplitude is constant inside each particle \cite{novotny}. However, these restrictions can be easily lifted by dividing the particles into sufficiently small dipolar subvolumes by following the well-established discrete dipole approximation \cite{purcell73,yaghjian80,yurkin07,novotny}. The formulas presented in this section are then simply written for a larger number of dipoles constituting the particles. The general forms of the Green's dyadic and the Langevin equations of motion remain unchanged, and consequently the mathematical derivation leading to the expressions \eqref{eq:landauer_formula}, \eqref{eq:tij_final} and \eqref{eq:tirad_final} for locally absorbed power is then identical with the derivation given here. 

\subsection{Quantum Langevin equations of motion}
\label{sec:equationofmotion}

We model the dipole dynamics of the polarization field in each particle $i$ by the classical oscillator model \cite{bornhuang,loudon70} accompanied by quantum Langevin dynamics \cite{ford65,ford88,dhar06} describing local thermal fluctuations. The equation of motion and its Fourier transform for the displacement coordinate $\bb{u}_i=\bb{p}_i/q$ corresponding to dipole moment $\bb{p}_i$ of dipole $i$ located at $\br_i$  are then
\begin{equation}
 m\ddot\bu_i(t) = -m\omega_i^2 \bu_i(t) +\xi_i(t) -m\gamma_i \dot{\bu}_i(t)  + q \bE_i(t) \label{eq:eom1}
\end{equation}
and
\begin{equation}
 -m\omega^2 \hat\bu_i(\omega) = -m\omega_i^2 \hat\bu_i(\omega) +\hat \xi_i(\omega) +im\gamma_i \omega \hat \bu_i(\omega)  + q \hat \bE_i(\omega), \label{eq:eom2}
\end{equation}
respectively. The local electric field $\bE_i$, which acts as a driving term in the equations of motion, is discussed below. The dipole mass $m$, resonance frequency $\omega_i$, Langevin friction constant $\gamma_i$ and charge $q$ are later incorporated into the definition of particle polarizability as discussed in Sec. \ref{sec:polarizabilities}.  

The friction term $m\gamma_i\dot{\bb{u}}_i$ represents the damping of dipole oscillations due to the coupling to a local heat bath and is responsible for dissipation, which appears as a non-zero imaginary part in the polarizability. Dissipation to the local heat bath is accompanied by thermal fluctuations described by the random force $\xi_i$, which turns the equation of motion into a stochastic differential equation. The relative magnitude of fluctuations and dissipation at each dipole site $i$ depends on the bath temperature $T_i$ through the fluctuation-dissipation relation presented in Sec. \ref{sec:noise}. For notational simplicity, we have assumed the friction to be memoryless and proportional to the instantaneous velocity with friction constant $\gamma$, representing Ohmic damping \cite{weiss}. For non-Ohmic bath, the derivation of the heat transfer rates proceeds similarly and the final results for heat transfer rates, derived in Sec. \ref{sec:lb_formula}, are given in a form valid for arbitrary damping. The Fourier transform in Eq. \eqref{eq:eom2} is defined, as usual, by $\hat f(\omega)=\int_{-\infty}^{\infty} dt e^{i\omega t} f(t)$ with the corresponding inverse transform $f(t)=\int_{-\infty}^{\infty} [d\omega/(2\pi)] e^{-i\omega t} \hat{f}(\omega)$. 

The Fourier-transformed local electric field $\hat{\bE}_i$ appearing in Eq. \eqref{eq:eom2} can be written following the discrete dipole approximation \cite{novotny}  as
\begin{equation}
 \hat{\bE}_i\pom= \Eenvhat(\br_i,\omega) + \sum_{j=1}^N \hat{\bE}_{ij}\pom. \label{eq:etot}
\end{equation}
Here $\Eenvhat(\br_i,\omega)$ is the stochastic thermal field originating from the environment and $\hat{\bE}_{ij}$ is the electric field due to dipole moment $\hat{\bb{p}}_j$, given in frequency domain by
\begin{equation}
 \hat{\bE}_{ij}(\omega) = \omega^2 \mu_0\gem_{ij}(\omega) \hat{\bp}_j (\omega). \label{eq:ekl}
\end{equation}
The electromagnetic Green's dyadic $\gem_{ij}$ appearing in Eq. \eqref{eq:ekl} is defined in Sec. \ref{sec:greens_dyadics}, where also the definition of the local Green's dyadic $\gem_{ii}$ accounting for the polarization field due to near-neighborhood and the radiation damping force is discussed. For Green's functions, only the frequency-domain representations are used in this paper so we omit their hats for brevity. 

Equation \eqref{eq:eom1} is semiclassical in the sense that we treat the displacements $\bb{u}_i$, noise $\xi_i$ and field $\Eenv$ as classical commuting variables and only include quantum effects by imposing quantum fluctuation-dissipation relations for the symmetrized correlators of $\xi_i$ and $\Eenv$ \cite{wang07}. We expect that a full quantum treatment of the mechanical degrees of freedom would give, after proper symmetrization of observables, identical results for heat transfer rates. This follows from the linearity of the equations of motion and is illustrated in Refs. \cite{wang07,saaskilahti13} for phonon heat transfer. 

The substitution of Eqs. \eqref{eq:etot} and \eqref{eq:ekl} to Eq. \eqref{eq:eom2} gives
\begin{equation}
 -m\omega^2 \hat{\bu}_i \pom =  -m\omega_i^2 \hat \bu_i\pom + \hat{\xi}_i\pom + im\gamma_i \omega \hat{\bu}_i\pom + \Eenvhat(\br_i,\omega) + q^2\omega^2\mu_0 \sum_{j=1}^N \gem_{ij}\pom \hat{\bu}_j\pom. \label{eq:eom_freq}
\end{equation}
Equation \eqref{eq:eom_freq} can be rearranged as
\begin{equation}
 - \sum_{j} \bb{A}_{ij} \hat{\bu}_j(\omega) = \hat{\xi}_i(\omega) + q\Eenvhat(\br_i,\omega) \label{eq:eom_akl}
\end{equation}
by defining an inverse propagator
\begin{equation}
 \bb{A}_{ij} = \left[m(\omega^2-\omega_i^2+i\gamma_i) \right]\delta_{ij}\bb{I}_{3\times 3}+ q^2\omega^2\mu_0 \gem_{ij},
\end{equation}
where $\unitdyadic$ is the $3\times 3$ unit matrix. The solution to Eq. \eqref{eq:eom_akl} can be written compactly in matrix form as
\begin{equation}
 \hat{\bu}(\omega) = -\bb{G}(\omega) \left[\hat{\xi}(\omega)+q\Eenvhat(\omega) \right], \label{eq:usol}
\end{equation}
where the dipole displacement Green's function $\bb{G}\pom=\bb{A}\pom^{-1}$ is
\begin{equation}
 \bb{G}(\omega) = \frac{1}{m(\omega^2 \mathbf{I}_{3N\times 3N}-\Omega^2)+q^2\omega^2\mu_0 \textrm{Re}[\gem(\omega)]+i\Gamma^{\textrm{bath}}(\omega)/2+i\Gamma^{\textrm{rad}}(\omega)/2}. \label{eq:g_expression1}
\end{equation}
Here we adopt a matrix notation where the dipole indices $i\in \{1,\dots,N\}$ and the spatial components $\alpha\in \{1,2,3\}$ are combined into a composite index resulting in matrices and vectors of size $3N\times 3N$ and $3N$, respectively. In the following, we will use an index notation where the subscript $ij$ ($i$) always refers to the $3\times 3$ matrix (3-component vector) corresponding to the notation used before Eq. \eqref{eq:usol}.

In Eq. \eqref{eq:g_expression1} we have additionally defined the block-diagonal resonance frequency matrix as  $\Omega=\textrm{diag}(\omega_1\bb{I}_{3\times 3},\omega_2\bb{I}_{3\times 3},\\ \dots,\omega_N\bb{I}_{3\times 3})$, the block-diagonal bath coupling matrix as $\Gamma^{\textrm{bath}} (\omega) = \textrm{diag}(2m \gamma_1 \omega\bb{I}_{3\times 3},\dots,2m\gamma_N \omega_N \bb{I}_{3\times 3})$,
and the radiation coupling function defined through the imaginary part of the electromagnetic interaction by\begin{equation}
 \Gamma^{\textrm{rad}}(\omega) = 2 q^2\omega^2\mu_0 \textrm{Im}[\gem(\omega)]. \label{eq:gammarad_def}
\end{equation}
In the denominator of Eq. \eqref{eq:g_expression1} and in all matrix-valued expressions appearing below, the scalar terms should be interpreted as being proportional to the unit matrix of size $3N\times3N$.

Equation \eqref{eq:g_expression1} shows that both the coupling to heat baths and the coupling to the radiation field produce broadening in the Green's function via the coupling functions $\Gamma^{\textrm{bath}}(\omega)$ and $\Gamma^{\textrm{rad}}(\omega)$. This broadening in the Green's function reflects dissipation that, along with the accompanying thermal fluctuations, enables energy transfer between dipoles with different bath temperatures.

\subsection{The electric Green's dyadic $\gem$}
\label{sec:greens_dyadics}
The electric Green's dyadic $\gem(\br,\br';\omega)$, which appears in Eq. \eqref{eq:ekl} through the definition $\gem_{ij}(\omega)\equiv \gem(\br_i,\br_j;\omega)$, is the Green's function of the nonhomogeneous Helmholtz equation for the electric field and defined by the equation \cite{novotny}
 \begin{equation}
 \nabla \times \nabla \times \gem(\bb{r},\br';\omega) - \kw^2 \epsenv(\br,\omega)\gem(\bb{r},\br';\omega)  =  \delta(\bb{r}-\br')\unitdyadic. \label{eq:gemdef}
\end{equation}
Here the wavenumber is $\kw=\omega/c$ defined in terms of frequency $\omega$ and the speed of light $c$ in vacuum. By definition, the Green's dyadic $\gem$ accounts for the scattering of the electromagnetic field from the inhomogeneities in the environment but not from the dipoles themselves. This effect is, however, automatically included in the formalism by the coupled dipole equations of motion. Since magnetic effects are neglected in this paper, we use the terms electric Green's dyadic and electromagnetic Green's dyadic interchangeably. 

The electric Green's dyadic can be separated into two parts as
\begin{equation}
 \gem(\bb{r},\bb{r}';\omega) = \gem^0(\bb{r},\bb{r}';\omega) + \gem^s(\bb{r},\bb{r}';\omega),
\end{equation}
where the free-space Green's dyadic $\gem^0(\bb{r},\bb{r}';\omega)$ is the particular solution of the inhomogeneous equation \eqref{eq:gemdef} with $\epsenv(\br,\omega)=\epsenv(\br',\omega)$, which in our case corresponds to $\epsenv(\br,\omega)=1$, \cite{novotny}
\begin{equation}
 \gem^0(\br,\br';\omega) = \left[\unitdyadic+ \frac{1}{\kw^2} \nabla \nabla^T  \right] \frac{e^{ik_0|\bb{r}-\bb{r}'|}}{4\pi|\bb{r}-\bb{r}'|}.
\end{equation}
The scattering part $\gem^s(\br,\br')$ is then the solution to the homogeneous equation that satisfies the proper boundary conditions due to inhomogeneities in $\epsenv(\br,\omega)$. The separation is necessary since the real part of the Green's dyadic $\gem^0$ diverges for $\br\to \br'$, so the local Green's dyadic $\gem_{ii}(\omega)$ appearing in Eq. \eqref{eq:ekl} is defined for each particle $i$ as \cite{yaghjian80,lakhtakia92,novotny}
\begin{equation}
 \mathbb{G}_{ii}(\omega)
  \approx \frac{1}{\Delta V_i} \left[\frac{1}{3} R^2 + \frac{i\kw}{6\pi} \Delta V_i - \frac{1}{3\kw^2} \right] \unitdyadic + \gem^s(\br_i,\br_i;\omega) \label{eq:mi_expr}
\end{equation}
where we have assumed each particle to be a sphere with radius $R_i$ and volume $\Delta V_i = 4\pi R_i^3/3$. We have also only included the lowest-order terms in $R$. The imaginary part of the term in braces in Eq. \eqref{eq:mi_expr} is the imaginary part of the local free-space Green's dyadic, $\textrm{Im}[\gem^0(\bb{r},\bb{r};\omega)]=i\omega/(6\pi c)  \unitdyadic$. The electric field produced by this local term is proportional to $\omega^3$ and produces the Abraham-Lorentz radiation damping \cite{jackson} proportional to the third time-derivative of the dipole moment in the equation of motion \eqref{eq:eom1}. The fluctuations accompanying the radiation damping are responsible for generating thermal radiation \cite{greffet10}. 

\subsection{Expressions for absorption and emission from Poynting's theorem}
\label{sec:poynting}
The most straightforward approach to evaluate the energy transfer rates between the dipoles is to first calculate the electric and magnetic fields $\bE$ and $\bb{H}$ due to the dipoles and to directly integrate the electromagnetic power flux obtained from the Poynting vector $\bb{S}=\bE \times \bb{H}$ over a surface enclosing the dipole under study. However, the calculation of the surface integrals is complicated and, in the present approach, one can simplify the problem by studying the heat transfer in terms of the dissipation and emission by the discrete dipoles.

To calculate the power absorbed by the dipole $i$, we shift momentarily back to time domain and apply Maxwell's equations $\nabla \times \bb{E} = - \mu_0 \partial \bb{H}/\partial t$ and $\nabla \times \bb{H}=\partial (\varepsilon_0 \bb{E}+\bb{P})/\partial t$, for non-magnetic media with polarization density $\bb{P}(\br,\omega)$ and no free charges, on the surface integral of the Poynting vector over the boundary $\partial V_i$ enclosing the particle volume $V_i$. By also applying the Gauss divergence theorem, we then get \cite{loudon70,rosa10}
\begin{alignat}{2}
 \int_{\partial V_i} \bb{S} \cdot d\bb{S} &= \int_{V_i} \nabla \cdot (\bb{E}\times \bb{H}) d\bb{r} \\
  &= - \int_{V_i}\frac{1}{2} \frac{\partial}{\partial t} \left[\varepsilon_0 \bb{E}^2 + \mu_0 \bb{H}^2 \right] d\bb{r} - \int_{V_i} \bb{E} \cdot \frac{\partial \bb{P}}{\partial t}   d\bb{r} \\
 &\approx  -  \int_{V_i} \frac{1}{2} \frac{\partial}{\partial t} \left[\varepsilon_0 \bb{E}^2 + \mu_0 \bb{H}^2 \right] d\bb{r} - \Delta V_i \bb{E}_i \cdot \frac{\dot{\bb{p}}_i}{\Delta V_i} \label{eq:poynting_4} \\
  &= - \frac{d }{d t} \int_{V_i} \frac{1}{2} \left[\varepsilon_0 \bb{E}^2 + \mu_0 \bb{H}^2 \right]  d\bb{r}    - \left[m\ddot{\bu}_i+m\omega_i^2\bu_i-\xi_i+m\gamma\dot{\bu}_i \right]\cdot \dot{\bu}_i  . \label{eq:poynting_3} 
\end{alignat}
In Eq. \eqref{eq:poynting_4}, we approximated the electric field inside the volume $V_i$ by the constant $\bE_i$ and similarly the polarization density by $\bb{P}(\br)= \bb{p}_i/\Delta V_i$. In moving to Eq. \eqref{eq:poynting_3}, we applied the equation of motion \eqref{eq:eom1} to solve for $\bE_i$. 

The expression \eqref{eq:poynting_3} is interpreted as the energy conservation law
\begin{equation}
 \int_{\partial V_i} \bb{S} \cdot d\bb{S} = - \frac{d }{d t} U_{i}^{\textrm{em}} - \frac{d}{dt} U^{\textrm{mech}}_i - Q_i^{\textrm{bath}}. \label{eq:energy_conservation}
\end{equation}
The first term on the right-hand side is the time derivative of the electromagnetic energy $
 U_{i}^{\textrm{em}} = \int_{V_i}  \frac{1}{2} \left[\varepsilon_0 \bb{E}^2 + \mu_0 \bb{H}^2 \right]d\bb{r}
$ stored in $V_i$, the second term is the rate of change of the mechanical energy $
U^{\textrm{mech}}_i=\frac{1}{2} m|\dot{\bb{u}}_i|^2+ \frac{1}{2} m\omega_i^2|\bb{u}_i|^2
$ stored in the oscillator $i$, and the third term 
\begin{equation}
 Q^{\textrm{bath}}_i=( m\gamma_i \dot{\bu}_i-\xi_i) \cdot\dot \bu_i \label{eq:qi_def}
\end{equation}
is the energy current to the Langevin heat bath. By thermal averaging (defined below in Sec. \ref{sec:noise}) and assuming steady state so that the total time-derivatives in Eq. \eqref{eq:energy_conservation} vanish, one gets $ \langle \int_{\partial V_i} \bb{S} \cdot d\bb{S} \rangle=-\langle Q^{\textrm{bath}}_i \rangle$, implying that the energy absorbed from the bath is equal to the emitted electromagnetic radiation.
 
\subsection{Statistical properties of bath noise and environment field}
\label{sec:noise}
In order to calculate the thermal average of the bath heat current \eqref{eq:qi_def}, we need to specify the statistical properties of the stochastic Langevin noise $\hat{\xi}$ and the thermal background field $\Eenvhat$ appearing in the solution \eqref{eq:usol} to the equation of motion \eqref{eq:eom2}. The Langevin noise has zero average $\langle \hat{\xi}_i \rangle= 0$ and its (symmetrized) autocorrelation function satisfies the quantum fluctuation-dissipation theorem \cite{dhar06}
\begin{alignat}{2}
 \langle \hat\xi_i(\omega) \hat\xi_j(\omega')^T \rangle &= 2\pi\delta(\omega+\omega') \hbar \Gamma^{\textrm{bath}}_i(\omega) \left[f_B(\omega,T_i) + \frac{1}{2} \right] \delta_{ij} ,\label{eq:xiixij}
\end{alignat}
where $T_i$ is the local bath temperature and the Bose-Einstein function is $f_B(\omega,T)=[\exp(\hbar\omega/k_B T)-1]^{-1}$. Here $\hbar$ is the reduced Planck's constant and $k_B$ is the Boltzmann constant. The delta function $\delta(\omega+\omega')$ reflects translational invariance in time and the Kronecker delta $\delta_{ij}$ ensures that the heat baths at sites $i$ and $j$ are uncorrelated, i.e., the baths are local. For the memoryless friction assumed in Eq. \eqref{eq:eom1}, the bath coupling function is $\Gamma^{\textrm{bath}}_i(\omega)=2m\gamma_i\omega \bb{I}_{3\times 3}$ as defined in Sec. \ref{sec:equationofmotion}. 

The thermal background field $\Eenvhat$ has zero average $\langle \Eenvhat \rangle=0$ and its symmetrized autocorrelation function satisfies the fluctuation-dissipation relation \cite{novotny}
\begin{alignat}{2}
  q^2 \langle \Eenvhat(\bb{r}_i,\omega) \Eenvhat(\bb{r}_j,\omega')^T \rangle   &=  2\pi \delta(\omega+\omega') \hbar \Gamma^{\textrm{rad}}_{ij}(\omega) \fbbg \label{eq:ebgebg1},
\end{alignat}
where $\Gamma^{\textrm{rad}}(\omega)$ was defined in Eq. \eqref{eq:gammarad_def}. The simultaneous presence of $\textrm{Im}[\gem_{ij}(\omega)]$ both in Eq. \eqref{eq:ebgebg1} and as a source of dissipation in the dipole displacement Green's function \eqref{eq:g_expression1} ensures the onset of thermal equilibrium when the dipole and environment bath temperatures are equal as discussed in Sec. \ref{sec:lb_formula}. 

We also note that the bath noises $\xi_i$ and the thermal background field $\Eenv$ are uncorrelated,
\begin{equation}
 \langle \hat \xi_i(\omega) \Eenvhat(\bb{r}_j;\omega')^T \rangle = 0 \qquad \forall i,j. \label{eq:xiiebg}
\end{equation}

\subsection{Energy exchange: Landauer-B\"uttiker formula}
\label{sec:lb_formula}
Having the solution \eqref{eq:usol} for the dipole displacements and noise correlations \eqref{eq:xiixij} and \eqref{eq:ebgebg1} available, we are equipped to calculate the thermal average of the energy flow \eqref{eq:qi_def} to each local heat bath. As shown in App. \ref{sec:appendix1}, the thermal average of the energy current flowing to the local heat bath at site $i$ can be written in the Landauer-B\"uttiker form
\begin{alignat}{2}
 \langle Q_i^{\textrm{bath}}\rangle &= \int_0^{\infty} \frac{d\omega}{2\pi}  \hbar \omega  \sum_{j=1}^N \ca{T}_{ij}\pom [f_B(\omega,T_j)-f_B(\omega,T_i)] \notag \\
  &\quad + \int_0^{\infty} \frac{d\omega}{2\pi}  \hbar \omega \ca{T}_{i,\textrm{rad}}\pom [f_B(\omega,\Tenv)-f_B(\omega,T_i)] ,\label{eq:landauer_formula}
\end{alignat}
where the energy transmission function between dipoles $i$ and $j$ is
\begin{equation}
  \ca{T}_{ij}\pom=\textrm{Tr}\left[\Gamma_i^{\textrm{bath}}\pom \bG_{ij}\pom \Gamma_j^{\textrm{bath}}\pom \bG_{ji}\pom^{\dagger}\right] \label{eq:tij}
\end{equation}
and between dipole $i$ and the radiation field 
\begin{equation}
 \ca{T}_{i,\textrm{rad}}\pom=\textrm{Tr}\left\{\Gamma_i^{\textrm{bath}}\pom \left[\bG\pom \Gamma^{\textrm{rad}}\pom \bG\pom^{\dagger}\right]_{ii} \right\}. \label{eq:tirad}
\end{equation}
Note that also here we use the index notation where $[\bG\pom \Gamma^{\textrm{rad}}\pom \bG\pom^{\dagger}]_{ii}$ is a $3\times3$ matrix describing the dyadic elements related to dipole $i$.

The first term of Eq. \eqref{eq:landauer_formula} accounts for the heat transfer between dipoles. The transmission function \eqref{eq:tij} is of the well-known Caroli form \cite{caroli71,datta}, giving the energy-dependent transmission function in terms of the Green's function $\bb{G}(\omega)$ and bath coupling functions $\Gamma^{\textrm{bath}}_i(\omega)$. In the context of electron \cite{buttiker92,datta} and phonon \cite{rego98,yamamoto06,saaskilahti13} transport, the bath coupling functions typically represent the coupling of the scattering region to semi-infinite leads, although they may also model coupling to local self-consistent voltage probes \cite{buttiker85} or heat baths \cite{bolsterli70} to account for inelastic scattering of electrons and phonons, respectively.

The second term of Eq. \eqref{eq:landauer_formula} consists of absorption due to the environment field at temperature $\Tenv$ and the radiation emitted by the dipole. The environment field compensates the energy loss due to radiation, ensuring that in the uniform temperature configuration, $T_{1}=\dots=T_N=\Tenv$, energy transfer to all local heat baths vanishes and the system is in thermal equilibrium. Equation \eqref{eq:landauer_formula} naturally implies that the total energy transfer to the radiation field is
\begin{equation}
 \langle Q^{\textrm{rad}} \rangle = \int_0^{\infty} \frac{d\omega}{2\pi} \sum_{i=1}^N  \hbar\omega \ca{T}_{i,\textrm{rad}}\pom [f_B(\omega,T_{i})-f_B(\omega,\Tenv)].
\end{equation}
As the transmission functions \eqref{eq:tij} and \eqref{eq:tirad} are symmetric, it is then easy to show that $\sum_{i=1}^N\langle Q_i^{\textrm{bath}} \rangle + \langle Q^{\textrm{rad}}\rangle=0$, showing that total energy is conserved. 

\subsection{Polarizabilities and the dielectric constant}
\label{sec:polarizabilities}
In this subsection we relate the parameters appearing in the dipole equation of motion \eqref{eq:eom1} and the transmission functions \eqref{eq:tij} and \eqref{eq:tirad} to the polarizabilities of the particles by studying the induced dipole moment $\langle \hat{\bb{p}_i}\pom \rangle$ when a single particle $i$ is placed in an external field $\hat{\bE}_0\pom$. For generality, we treat the particle polarizabilities as tensors, since the final expressions for the transmission functions are valid also for anisotropic media. Anisotropy would simply appear in the equation of motion \eqref{eq:eom1} through direction-dependent parameters and, in the case of a non-principal coordinate system, as coupling between the different Cartesian components of the dipole displacement vector.

For completeness and due to the several widely used definitions for the local polarizability tensor, we specifically define the bare, Clausius-Mossotti and effective polarizabilities through relations
\begin{subequations}
\begin{align}
\langle \hat{\bp}_i(\omega) \rangle&= \varepsilon_0 \alpha_0^i(\omega) \langle \hat{\bb{E}}_i(\omega) \rangle, \label{def:alpha0} \\
 \langle\hat{\bp}_i(\omega) \rangle &= \varepsilon_0 \alpha_{\textrm{CM}}^i(\omega) \left[\hat{\bb{E}}_0(\br_i,\omega)+\langle \hat{\bE}_{\textrm{pol},i}(\omega)\rangle\right], \\
  \langle\hat{\bp}_i(\omega) \rangle &= \varepsilon_0 \alpha_{\textrm{eff}}^i(\omega) \hat{\bb{E}}_0(\br_i,\omega),
\end{align}
\end{subequations}
where the bare polarization $\alpha_0^i(\omega)$ relates the dipole moment and the total macroscopic electric field $\hat{\bE}_i\pom= \hat{\bb{E}}_0(\br_i,\omega)+\omega^2 \mu_0 \gem_{ii}\pom \hat{\bb{p}}_i\pom$, the Clausius-Mossotti polarizability $\alpha_{\textrm{CM}}^i\pom$ relates the dipole moment to the sum of $\hat{\bE}_i\pom$ and the polarization field $\hat{\bE}_{\textrm{pol},i}(\omega)= \hat{\bb{p}}_i\pom /(3\varepsilon_0 \Delta V_i)$, the latter modifying the local field seen by the dipole, and the effective polarizability $\alpha_{\textrm{eff}}^i(\omega)$ relates the dipole moment directly to the external field.

By solving for $\langle \hat{\bu}_i\pom \rangle$ from the equation of motion \eqref{eq:eom1} for a single dipole in an external field $\hat{\bb{E}}_0(\omega)$, one gets
\begin{subequations}
 \begin{align}
 \alpha_0^i(\omega) &= - \frac{q^2}{\varepsilon_0} \frac{1}{m(\omega^2-\omega_i^2+i\gamma_i)} \unitdyadic, \label{eq:alpha0_expr} \\
\alpha_{\textrm{CM}}^i(\omega) &= - \frac{q^2}{\varepsilon_0} \frac{1}{m(\omega^2-\omega_i^2+i\gamma_i)-q^2/(3\varepsilon_0\Delta V_i)}\unitdyadic, \label{eq:alpha_cm_expr}  \\
 \alpha_{\textrm{eff}}^i(\omega) &= - \frac{q^2}{\varepsilon_0} \frac{1}{m(\omega^2-\omega_i^2+i\gamma_i)\unitdyadic+q^2\omega^2\mu_0 \mathbb{G}_{ii}(\omega)}. \label{eq:alpha_eff_expr}
\end{align}
\end{subequations}
It is easy to see that the polarizabilities are related by
\begin{subequations}
\begin{align}
 \alpha_{\textrm{CM}}^i(\omega)^{-1} &= \alpha_0^i(\omega)^{-1} + \frac{1}{3\Delta V_i} \unitdyadic ,\label{eq:alphacm_alpha0} \\
  \alpha_{\textrm{eff}}^i(\omega)^{-1} &= \alpha_{0}^i(\omega)^{-1} - \kw^2 \mathbb{G}_{ii}(\omega). \label{eq:alphaeff_alphacm}
\end{align}
\end{subequations}
Note that the definition of the effective polarizability depends on the environment through the local Green's dyadic $\gem_{ii}(\omega)$ [Eq. \eqref{eq:gemdef}]. Definition based on the effective polarizability in free-space is obtained by replacing $\gem_{ii}(\omega)$ in Eq. \eqref{eq:alpha_eff_expr} by the free-space Green's dyadic $\gem_{ii}^0(\omega)$ [Eq. \eqref{eq:mi_expr}].

The polarizabilities are directly related to the known bulk dielectric constants $\varepsilon_i(\omega)=1+\chi_i(\omega)$ of the particles through $\hat{\bb{P}}(\bb{r}_i,\omega)=\varepsilon_0 \chi_i(\omega) \hat{\bb{E}}(\bb{r}_i,\omega)$ and $\hat{\bb{P}}(\bb{r}_i,\omega)=\hat{\bb{p}}_i(\omega)/\Delta V_i$. Using the definition \eqref{def:alpha0} of the bare polarizability, we see that $\chi_i(\omega)\unitdyadic=\alpha_0^i(\omega)/\Delta V_i$. Using also Eqs.  \eqref{eq:alphacm_alpha0}-\eqref{eq:alphaeff_alphacm}, one gets
\begin{subequations}
\begin{align}
 \alpha_0^i(\omega) &=\Delta V_i [\varepsilon_i(\omega)-1]\unitdyadic \\
   \alpha_{\textrm{CM}}^i(\omega) &= 3\Delta V_i \frac{\varepsilon_i(\omega)-1}{\varepsilon_i(\omega)+2} \unitdyadic, \label{eq:alphacm_epsilon} \\
     \alpha_{\textrm{eff}}^i(\omega) &=3 \Delta V_i \frac{\varepsilon_i(\omega)-1}{\varepsilon_i(\omega)+2} \times \left\{ {\bb{I}_{3\times 3}- \frac{3\kw^2[\varepsilon_i(\omega)-1]}{\varepsilon_i(\omega)+2}} \mathbb{M}_i \right\}^{-1}. \label{eq:alphaeff_final}
\end{align}
\end{subequations}
Equation \eqref{eq:alphacm_epsilon} is the well-known Clausius-Mossotti relation (or Lorentz-Lorenz relation in optics) connecting the local polarizability to the dielectric constant \cite{bornwolf}. In Eq. \eqref{eq:alphaeff_final}, we have defined the dyadic $\mathbb{M}_i$ that excludes the polarization dyadic from the local Green's function $\gem_{ii}\pom$ [Eq. \eqref{eq:mi_expr}] as
\begin{alignat}{2}
 \mathbb{M}_i \pom 
  &\approx \left[\frac{1}{3} R^2 + \frac{i\kw}{6\pi} \Delta V_i \right] \unitdyadic + \Delta V_i \gem^s(\br_i,\br_i;\omega).
\end{alignat}
Definition \eqref{eq:alphaeff_final} arises in the discrete dipole approximation as the unique definition of the local polarizability that ensures the equivalence of the method of coupled dipoles (which solves the dipole moments) with the method of moments (which solves the electric fields) \cite{lakhtakia92,novotny}. 

As a side note, we note that our formalism inherently includes the modified fluctuation-dissipation theorem derived by Manjavacas and Garc\'ia de Abajo \cite{manjavacas10,manjavacas12}. The fluctuation-dissipation theorem for dipole moments can be derived from our formalism by noting that for a single dipole coupled to a thermal bath at temperature $T_i$ and located in empty space with no environment field, the expectation value of the dipole moment correlation function can be written as
 \begin{equation}
  \langle \bb{p}_i(\omega) \bb{p}_i(\omega')^T \rangle =4\pi \hbar\varepsilon_0 \tilde{\chi}_i(\omega)\left[f_B(\omega,T_i)+\frac{1}{2}\right] \delta(\omega+\omega') \label{eq:pp_corr},
 \end{equation}
where the reduced susceptibility tensor $\tilde{\chi}_i$ (note the difference to the bulk electric susceptibility $\chi_i$) is defined in terms of the effective polarizability $\alpha^i_{\textrm{eff}}(\omega)$ as 
\begin{equation}
 \tilde{\chi}_i(\omega) = \textrm{Im}[\alpha^i_{\textrm{eff}}(\omega)] - \frac{\kw^3}{6\pi} \alpha_{\textrm{eff}}^i(\omega) \alpha_{\textrm{eff}}^i(\omega)^{\dagger}. \label{eq:gammabath_chi}
\end{equation}
We do not, however, need to impose Eq. \eqref{eq:pp_corr} for the dipole moment fluctuations, since fluctuations are naturally included in the formalism by the coupling to Langevin baths.

\subsection{Transmission functions in terms of polarizabilities}
\label{sec:transmission_polarizability}

Since all three definitions \eqref{eq:alpha0_expr}, \eqref{eq:alpha_cm_expr} and \eqref{eq:alpha_eff_expr} of the polarizability can be directly related to the dielectric constant, the choice of which polarizability to use is a matter of taste. We choose to express the transmission functions \eqref{eq:tij} and \eqref{eq:tirad} in terms of the Clausius-Mossotti polarizability. We first note that the term $m\gamma_i \omega$, reflecting the strength of coupling to the local heat bath,  can be solved from the definition \eqref{eq:alpha_cm_expr} of the Clausius-Mossotti polarizability to get the bath coupling function 
\begin{alignat}{2}
 \Gamma_i^{\textrm{bath}}(\omega) &= 2m\gamma_i \omega \unitdyadic \\
 &=\frac{2q^2}{\varepsilon_0} [\alpha^i_{\textrm{CM}}(\omega)]^{-1} \textrm{Im}[\alpha^i_{\textrm{CM}}(\omega)][\alpha^i_{\textrm{CM}}(\omega)^{\dagger}]^{-1}. \label{eq:gammabath_cm}
\end{alignat}
Similarly we can express the dipole displacement Green's function \eqref{eq:g_expression1} in terms of the block-diagonal polarizability matrix $\alpha_{\textrm{CM}} (\omega) = \textrm{diag}[\alpha^1_{\textrm{CM}}(\omega),\dots,\alpha^N_{\textrm{CM}}(\omega)]$ as
\begin{alignat}{2}
 \bG \pom &= -\frac{\varepsilon_0}{q^2} \alpha_{\textrm{CM}}(\omega) \frac{1}{1-\kw^2 \gem(\omega)\alpha_{\textrm{CM}}(\omega)}. \label{eq:galphab}
\end{alignat}
Note that the unknown microscopic parameters $m$, $\omega_i$ and $\gamma_i$ appearing in the dipole equation of motion $\eqref{eq:eom1}$ and in the dipole displacement Green's function \eqref{eq:g_expression1} have now been fully absorbed into the particle polarizabilities, which can in turn be inferred from the dielectric constants using Eq. \eqref{eq:alphacm_epsilon}. 

Substituting Eq. \eqref{eq:gammabath_cm}, its complex conjugate and Eq. \eqref{eq:galphab} to Eq. \eqref{eq:tij} for $\ca{T}_{ij}(\omega)$, one gets after straightforward algebraic manipulations
\begin{equation}
   \ca{T}_{ij}(\omega) = 4 \kw^4 \textrm{Tr} \left[ \textrm{Im}[\alpha^i_{\textrm{CM}}(\omega)] \gemfull_{ij}\pom \textrm{Im}[\alpha^j_{\textrm{CM}}(\omega)] \gemfull_{ji}(\omega)^{\dagger}\right]. \label{eq:tij_final}
\end{equation}
Here the electromagnetic Green's dyadic $\gemfull(\omega)$ expressed in terms of the Clausius-Mossotti polarizabilities of the particles has been defined as 
\begin{alignat}{2}
 \gemfull(\omega) &= \frac{1}{\kw^2} \left[\frac{1}{1-\kw^2 \gem(\omega)\alpha_{\textrm{CM}}(\omega)}\right] \alpha_{\textrm{CM}}(\omega)^{-1} \\
  &\equiv \frac{1}{\kw^2} \alpha_{\textrm{CM}}(\omega)^{-1} + \underbrace{ \left[\frac{1}{1-\kw^2 \gem(\omega)\alpha_{\textrm{CM}}(\omega)} \right] \gem(\omega)}_{\tildegemfull}. \label{eq:gemmb_cm_app}
\end{alignat}
The first term of Eq. \eqref{eq:gemmb_cm_app} is local and does not contribute to dipole-dipole energy transfer. The second term of Eq. \eqref{eq:gemmb_cm_app}, denoted by $\tildegemfull$, is non-local and therefore responsible for dipole-dipole energy transfer. It is easy to see that this part satisfies the equation
\begin{equation}
 \tildegemfull (\omega) = \gem(\omega) + \kw^2 \gem(\omega) \alpha_{\textrm{CM}}(\omega) \tildegemfull(\omega).\label{eq:gem_dyson}
\end{equation}
Comparing with the well-known Dyson equation satisfied by the electromagnetic Green's dyadic in an inhomogeneous environment \cite{martin95} and noting that $\gem$ is the Green's dyadic unperturbed by the dipoles, one notices that $\tildegemfull$ can be interpreted as the electromagnetic Green's dyadic that fully incorporates the scattering caused by the dipoles. Equation \eqref{eq:gem_dyson} arises also from the combination of Eqs. \eqref{eq:etot} and \eqref{eq:ekl}, if one divides the dipole moment into the induced and fluctuating parts and calculates the electric field due to the fluctuating parts \cite{benabdallah11}. Here Eq. \eqref{eq:gemmb_cm_app} arises as a convenient definition that allows us to express the transmission function \eqref{eq:tij_final} in terms of purely optical quantities.

The radiation transmission function \eqref{eq:tirad} can be similarly written in terms of the polarizabilities and the Clausius-Mossotti Green's dyadic as
\begin{alignat}{2}
 \ca{T}_{i,\textrm{rad}}(\omega) 
 &= 4 \kw^6 \sum_{j,k=1}^N \textrm{Tr}\left\{   \textrm{Im}[\alpha^i_{\textrm{CM}}(\omega)] \gemfull_{ij}(\omega) \alpha_{\textrm{CM}}^j(\omega) \textrm{Im}[\gem_{jk}(\omega)] \alpha^k_{\textrm{CM}}(\omega)^* \gemfull_{ki}(\omega)^{\dagger} \right\}. \label{eq:tirad_final}
\end{alignat}
Note that also the first, local term of Eq. \eqref{eq:gemmb_cm_app} contributes to the sum through the local terms $\gemfull_{ii}$. It is straightforward to derive similar expressions for $\ca{T}_{ij}(\omega)$ and $\ca{T}_{i,\textrm{rad}}(\omega)$ also in terms of the bare and effective polarizabilities, but since the final results are anyway equivalent, only Eqs. \eqref{eq:tij_final} and \eqref{eq:tirad_final} are presented here. 

Equations \eqref{eq:tij_final} and \eqref{eq:tirad_final} for the transmission functions have been derived also from fluctuational electrodynamics (FED) \cite{benabdallah11,messina13}. The main difference between FED and Langevin dynamics (LD) is that in FED, thermal dipole fluctuations are described by the dipole fluctuation-dissipation theorem \eqref{eq:pp_corr}. Accompanied with self-consistent equations for the electric fields and the induced parts of the dipole moments, the locally dissipated energy at each dipole site can be calculated from the product of the local electric field and the induced dipole current. In Langevin dynamics, on the other hand, dipolar fluctuations are described by stochastic forces in the microscopic dipole equation of motion \eqref{eq:eom1}, which separately describes all driving and dissipative forces acting on the dipole displacements. With the help of Poynting's theorem, we showed that the calculation of the locally absorbed power from LD does not require solving for the electric fields: only the dipole moments \eqref{eq:usol} and the FDTs \eqref{eq:xiixij} and \eqref{eq:ebgebg1} are required [Eq. \eqref{eq:qi_def}]. 

Langevin dynamics has also other advantages compared to FED. First, we were able to show that the dipole-dipole energy transmission function \eqref{eq:landauer_formula} has exactly the same Caroli form \cite{caroli71} as in electronic \cite{datta} and phononic \cite{yamamoto06} transport. Our formalism therefore presents an appealing unification of photonic energy transfer with the electronic and phononic theories, promoting the development of coupled energy transfer models covering all three carriers. Second, our methodology offers a simple, alternative derivation of the modified fluctuation-dissipation theorem \eqref{eq:pp_corr}, in which the Abraham-Lorentz radiation damping [Eq. \eqref{eq:mi_expr}] is subtracted from the imaginary part of the effective polarizability \eqref{eq:alphaeff_final} in the fluctuation-dissipation theorem to avoid non-physical fluctuations in non-absorbing particles \cite{manjavacas12}. In FED, this subtraction is not inherently built in the theory but has to be derived separately. Finally, Langevin formalism does not require artificial separation between the ''fluctuating'' and ''induced'' parts of the dipole moment as in FED: only the total dipole moment appears in our formalism.

\section{Results and discussion}
\label{sec:results}

The well-known Purcell effect \cite{purcell46} symbolizes the strong position dependence of the spontaneous emission rate of an atom placed in an optical cavity. The same effect has also been reported for dipole-dipole interactions in a resonant cavity \cite{kobayashi95,agarwal98,hopmeier99}, but the details of the thermal energy transfer and the perturbations caused by the dipoles themselves have not previously been studied in cavities including several dipoles. As an example of the developed formalism, we investigate in Sec. \ref{sec:results_cavity} how cavity modes modify the energy exchange rate between SiC nanoparticles in a microcavity. In Sec. \ref{sec:results_surface}, we also investigate the modification of the interparticle heat transfer rate above a SiC surface, which can support coupled surface modes of phonons and photons known as surface phonon polaritons \cite{joulain05}. 

In both configurations, the nanoparticle radii are set to $R=250$ nm and the dielectric constant of SiC is modeled by the Lorentz model \cite{spitzer59,mulet01}
\begin{equation}
  \varepsilon(\omega) = \varepsilon_{\infty} \left(1+ \frac{\omega_L^2-\omega_T^2}{\omega_T^2-\omega^2-i\Gamma\omega} \right)
\end{equation}
where $\varepsilon_{\infty}=6.7$, $\hbar \omega_L=120.2$ meV, $\hbar\omega_T=98.6$ meV and $\hbar\Gamma=590.6$ $\mu$eV. The imaginary parts of the particle polarizabilities $\alpha_{\textrm{CM}}(\omega)$ [Eq. \eqref{eq:alphacm_epsilon}] peak strongly at the resonance energy $\hbar \omega_{r}=115.6$ meV corresponding to the condition $\textrm{Re}[\varepsilon(\omega_r)]=-2$. The strong absorptance peak at this energy dominates the heat transfer rates at room temperature, and the associated resonant mode is sometimes called the Fr\"ohlich mode \cite{bohren}. 

\subsection{Heat transfer between SiC nanoparticles in a resonant cavity}
\label{sec:results_cavity}
\begin{figure}[t]
 \begin{center}
 \includegraphics[width=8.6cm]{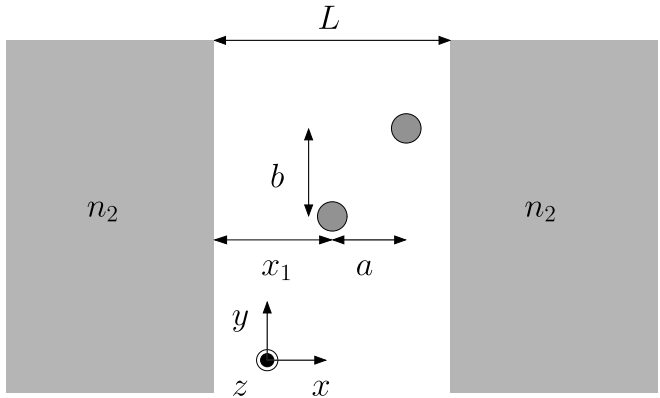}
 \caption{Schematic illustration of the cavity geometry studied in Sec. \ref{sec:results_cavity}. Particle $1$ is located at the center of a planar cavity ($x_1=L/2$) and the location of the particle $2$ is varied.}
 \label{fig:cavity_geom}
 \end{center}
\end{figure}

\begin{figure}[t]
 \begin{center}
 \includegraphics[width=.6\columnwidth]{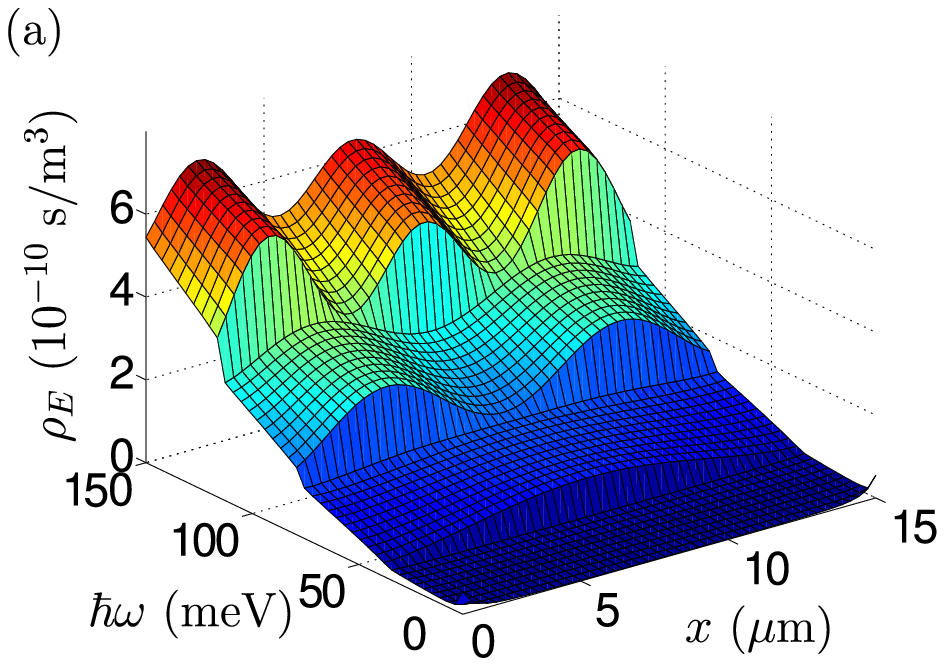}
 \includegraphics[width=.3\columnwidth]{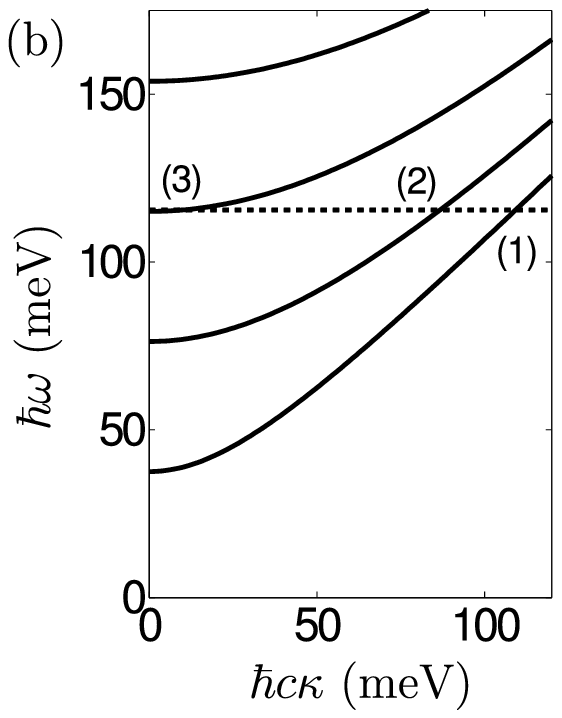}
 \caption{(Color online) (a) The electric density of states $\rho_E(\omega)=\omega \textrm{Im}\{\textrm{Tr}[\gem(x,x;\omega)]\}/(\pi c^2)$ in an empty cavity of length $L=16$ $\mu$m as a function of distance $x$ from the cavity wall and energy $\hbar \omega$, with arbitrary in-plane positions $y_1=y_2$ and $z_1=z_2$. The confined modes appear as position-dependent densities of states above threshold energies $E_n=\hbar c \beta_n$. (b) The in-plane dispersion of the confined electromagnetic modes in the cavity. The resonance energy $\hbar \omega_{r}=115.6$ meV of a SiC nanoparticle, shown as the dashed line, can couple to the third mode at $\hbar c \kappa_3\approx 11$ meV [crossing denoted by (3)], to the second mode at $\hbar c \kappa_2 \approx 87$ meV (2) and to the first mode at $\hbar c \kappa_1 \approx 109$ meV (1). The corresponding in-plane wavelengths $\lambda_n=2\pi/ \kappa_n$ are $\lambda_1\approx 11$ $\mu$m, $\lambda_2\approx 14$ $\mu$m and $\lambda_3=116$ $\mu$m for (1), (2) and (3), respectively.}
 \label{fig:rho}
 \end{center}
\end{figure}

In this section, the nanoparticles are assumed to be located in a planar cavity of length $L$, illustrated in Fig. \ref{fig:cavity_geom}. The cavity Green's dyadic $\gem(\omega)$ is calculated in the presence of the cavity walls occupying the half-spaces $x<0$ and $x>L$ \cite{tomas95}. The space between the walls is vacuum corresponding to the refractive index $n_1=1$. The refractive index in the two half-spaces is chosen to be $n_2=2+20i$, which corresponds to the reflection coefficient $R\approx 0.986$ and phase shift $\phi=5.67^{\circ}$ at normal incidence. The environment dielectric function is then $\epsenv(\br,\omega)=(n_2)^2$ for $x<0$ and $x>L$ and unity otherwise. We decompose the wavevector $\bb{k}$ into longitudinal and in-plane parts as $\bb{k}=(\beta,\vec \kappa)$. As the cavity walls are very reflective, modes satisfying the constructive interference condition
\begin{equation}
 2\beta_n L + 2\phi = 2\pi n, \quad n \in \mathbb{Z},
\end{equation}
show up in the local electromagnetic density of states in Fig. \ref{fig:rho}(a), where the electric density of states $\rho_E(\omega)=\omega\textrm{Im}\{\textrm{Tr}[\gem(x,x;\omega)]\}/(\pi c^2) $ \cite{joulain05} in a cavity of length $L=16$ $\mu$m is shown. The threshold energies of the cavity modes are $E_n=\hbar c \beta_n$, the smallest three of which are visible at energies $E_1=37.6$ meV, $E_2=76.3$ meV, and $E_3=115.1$ meV. For frequency $\omega$, the corresponding in-plane wavevectors are $\kappa=\sqrt{(\omega/c)^2-\beta_n^2}$, whose dispersion is shown in Fig. \ref{fig:rho}(b). The Fr\"ohlich resonance energy of a SiC nanoparticle, shown as the dashed line, can couple to each of the three modes at different in-plane wavevectors $\kappa$.

\begin{figure}[t]
 \begin{center}
 \includegraphics[width=.49\columnwidth]{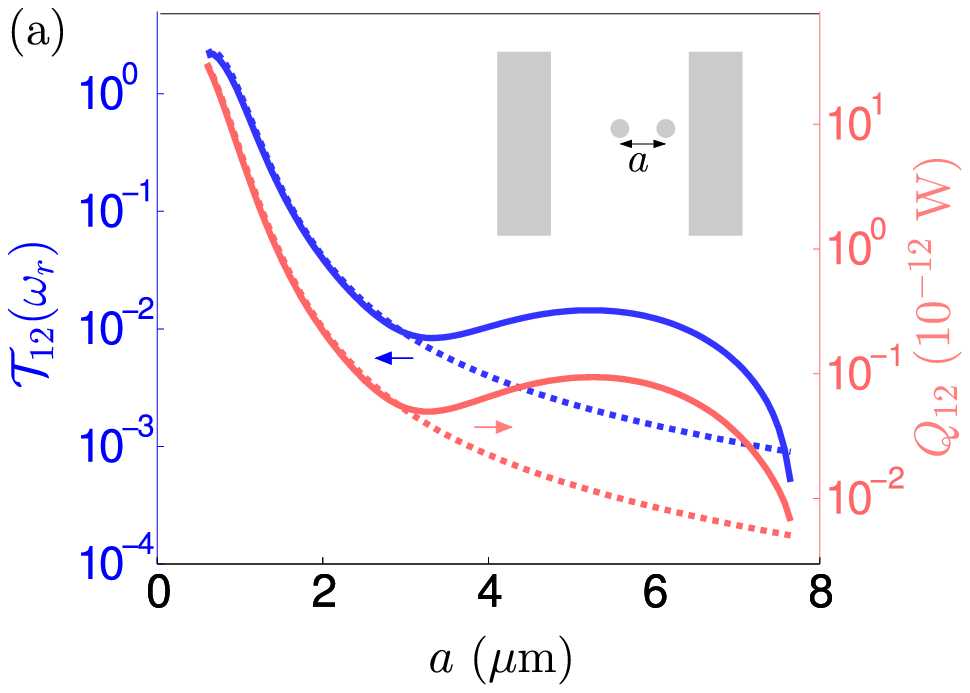}
 \includegraphics[width=.49\columnwidth]{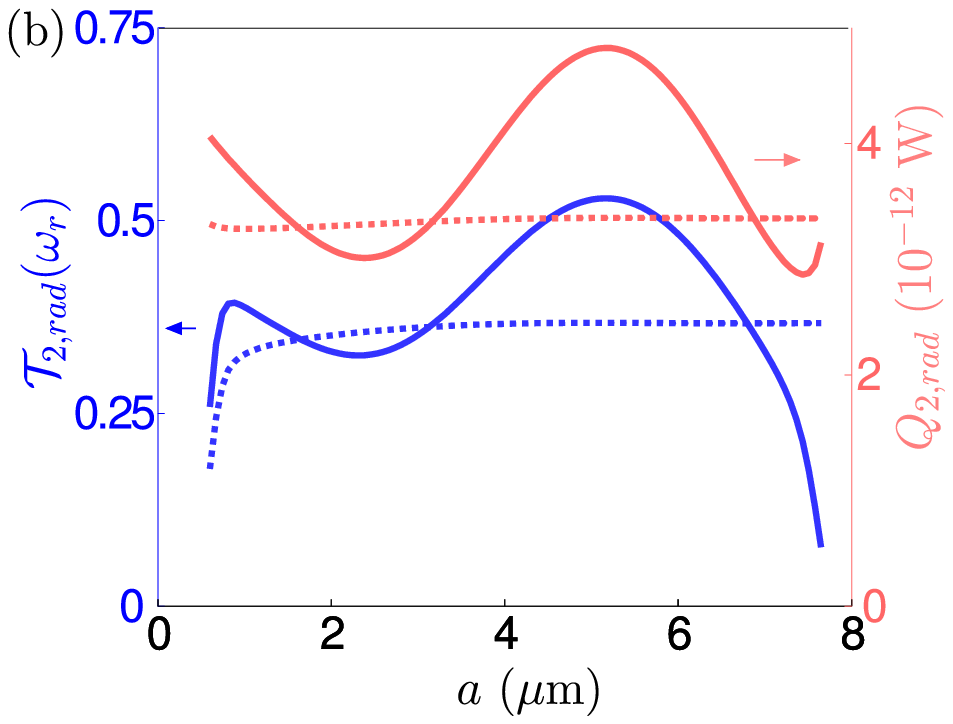}
 \caption{(Color online) (a) Transmission function $\ca{T}_{12}(\omega_{r})$ (solid blue or dark gray, left axis) and the interparticle heat current $Q_{12}$ (solid red or light gray, right axis) as a function of SiC nanoparticle separation $a$ with in-plane separation $b=0$. One particle is fixed at the center of the cavity, $x_1=L/2$, as illustrated in the inset. The vertical axes are logarithmic and the free space transmission functions and heat currents are shown as dashed lines.  In the transmission function, the frequency $\omega$ is chosen to correspond to the nanoparticle mode resonance $\hbar \omega_{r}=115.6$ meV, where the particles can strongly couple to the third standing mode [point (3) in Fig. \ref{fig:rho}(b)]. For the heat current, the particle temperatures are $T_1=310$ K and $T_2=300$ K. The cavity length is $L=16$ $\mu$m and the nanoparticle radius $R=250$ nm. (b) Transmission function $\ca{T}_{2,\textrm{rad}}$ and the heat current $Q_{2,\textrm{rad}}$ as a function of $a$ for cavity (solid) and free space (dashed). For the heat current, the environment temperature is $\Tenv=310$ K and the particle temperature is $T_2=300$ K. In contrast to (a), the vertical axes are linear.}
 \label{fig:T12_a}
 \end{center}
\end{figure}

Figure \ref{fig:T12_a}(a) shows the transmission functions $\ca{T}_{12}(\omega_r)$ and the interparticle heat current $Q_{12}$ as a function of particle distance $a$ for two SiC nanoparticles in a cavity. One of the particles is fixed at the center of the cavity at $x_1=L/2$ and the in-plane separation is set to $b=0$ as shown in the inset. The cavity length is $L=16$ $\mu$m and the frequency in the transmission function is set to correspond to the nanoparticle Fr\"ohlich resonance $\hbar \omega_r=115.6$ meV. In the heat current calculation, the particle temperatures are set to $T_1=310$ K and $T_2=300$ K. The heat current closely follows the trends of the transmission function due to the strong absorptance peak at the Fr\"ohlich mode frequency $\omega_r$. For small $a$ (but $a\gg R$) the interparticle heat current $\ca{Q}_{12}$ diverges as $1/a^{6}$ due to the near-field divergence of the free-space Green's dyadic \cite{domingues05}. As the separation $a$ increases, the interparticle heat current quickly starts to decrease. Due to the coupling to the cavity modes, the transmission function and the heat current have, however, another local maximum at $a\approx 5$ $\mu m$, corresponding to the amplitude maximum of the third standing mode at $x\approx 5L/6$ in Fig. \ref{fig:rho}(a) and depicted by point (3) in Fig. \ref{fig:rho}(b). At the amplitude maximum, the interparticle transmission is enhanced ten-fold compared to the free-space value (dashed line). 

The transmission function $\ca{T}_{2,\textrm{rad}}(\omega_r)$ and the particle-environment heat current $Q_{2,\textrm{rad}}$ are plotted in Fig. \ref{fig:T12_a}(b). In the heat current calculation, we set the environment temperature to $\Tenv=310$ K. Coupling to the radiation field depends on the particle position due to the position-dependent coupling to the third cavity mode and results in the well-known Purcell effect. At $a\approx 5$ $\mu$m, where the interparticle transmission $\ca{T}_{12}$ has a local maximum as a function of particle $2$ position, the transmission $\ca{T}_{2,\textrm{rad}}$ also has a local maximum and the value exceeds the interparticle transmission nearly by two orders of magnitude. This implies that if the temperature of the nanoparticle 1 was increased above $\Tenv$, the steady-state temperature $T_2$ of the particle $2$ would be dominated by the environment temperature $\Tenv$, unless $T_1 \gg \Tenv$. The latter condition could be achieved either by strong heating of particle 1 or by the cooling of the environment.

\begin{figure}[t]
 \begin{center}
 \includegraphics[width=.49\columnwidth]{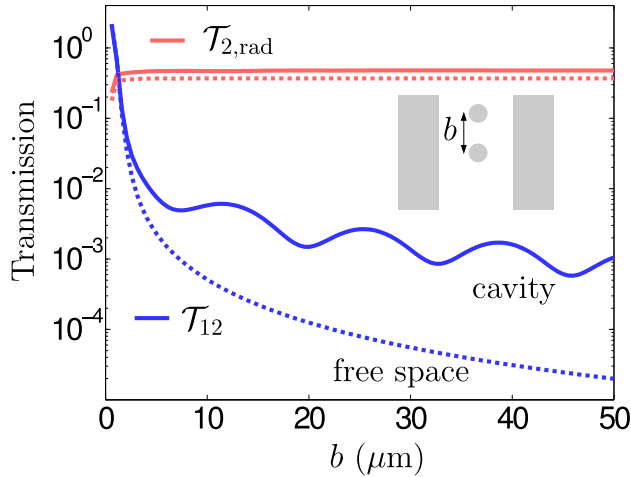}
 \caption{(Color online) Transmission functions $\ca{T}_{12}(\omega_r)$ and $\ca{T}_{2,\textrm{rad}}(\omega_r)$ as a function of in-plane separation $b$ for $a=0$, with cavity length $L=16$ $\mu$m and frequency $\hbar \omega_r=115.6$ meV. The vertical axis is logarithmic and the geometry is illustrated in the inset. The dashed lines show the transmission functions in free space.}
 \label{fig:T12_b}
 \end{center}
\end{figure}

Figure \ref{fig:T12_b} shows the transmission functions $\ca{T}_{12}(\omega_r)$ and $\ca{T}_{2,\textrm{rad}}(\omega_r)$ as a function of in-plane separation $b$ of the SiC nanoparticles. Apart from the near-field, the interparticle transmission function $\ca{T}_{12}$ is again visibly larger (on the logarithmic scale) in the cavity than in free space. The transmission function oscillates as a function of $b$ due to the interference of three propagating cavity modes. The oscillation can be shown to be an interference effect by evaluating $\ca{T}_{12}$ in a narrow cavity, where only one of cavity modes can be excited at $\hbar \omega=115.6$ meV, and noting that the oscillations vanish (not shown). The enhanced interparticle energy transfer demonstrated in Figs. \ref{fig:T12_a}(a) and \ref{fig:T12_b} could have applications, for example, in enhancing the spatial and temporal control of plasmonic heating in metal nanoparticle collections\cite{cao07,adleman09,yannopapas13}. 

\subsection{Heat transfer between SiC nanoparticles above a SiC surface}
\label{sec:results_surface}

\begin{figure}[t]
 \begin{center}
 \includegraphics[width=.49\columnwidth]{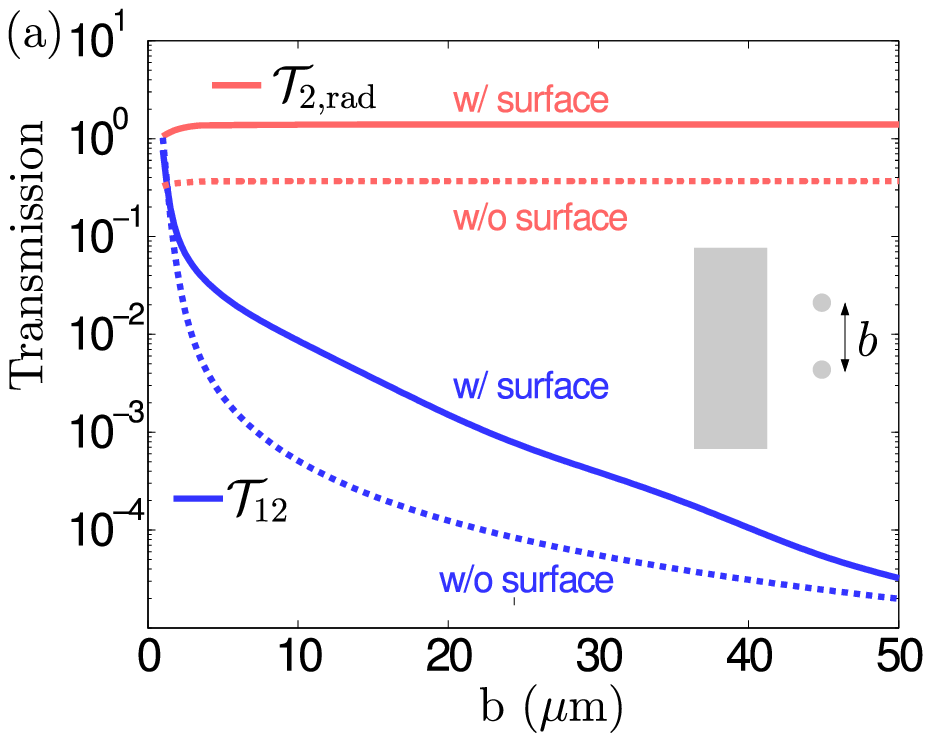}
 \includegraphics[width=.49\columnwidth]{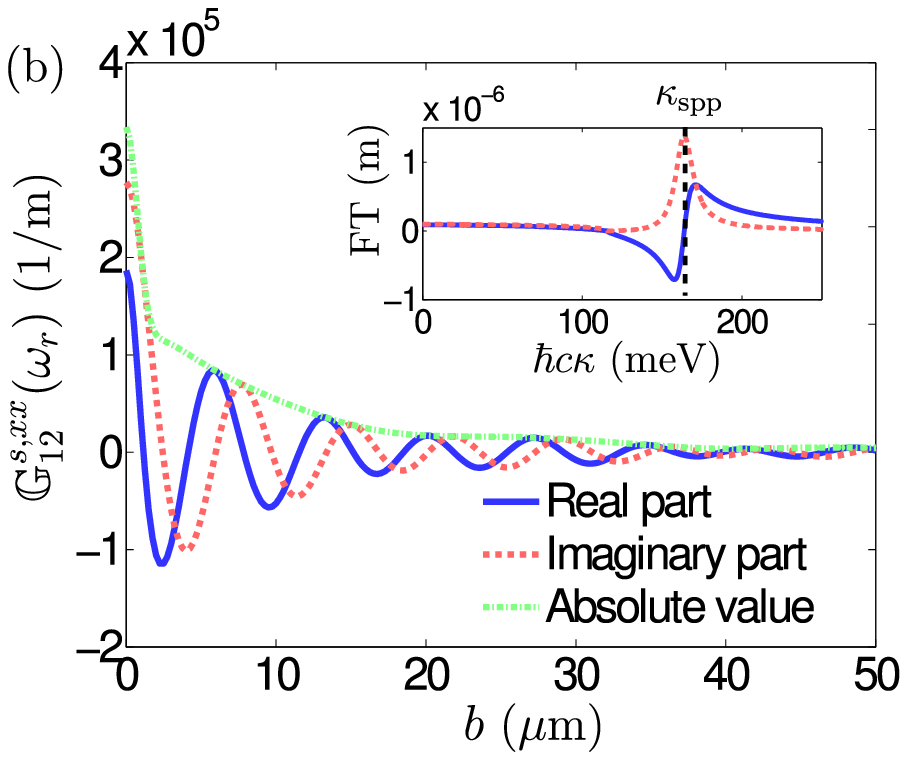}
 \caption{(Color online) (a) Transmission functions $\ca{T}_{12}(\omega_r)$ (blue or dark gray) and $\ca{T}_{2,\textrm{rad}}(\omega_r)$ (red or light gray) at $\hbar \omega_r=115.6$ meV as a function of in-plane nanoparticle separation $b$, with (solid) and without (dashed) a SiC surface. The distance from the surface is $1$ $\mu$m. Both interparticle transmission $\ca{T}_{12}$ and the radiation coupling are strongly enhanced compared to the free-space value by the coupling to surface phonon polaritons. The geometry is illustrated in the inset. (b) The real and imaginary parts and the absolute value of the $xx$-component of the interparticle Green's dyadic $\gem^{s}_{12}(\omega)$ as a function of $b$. Both real and imaginary parts oscillate at the surface phonon polariton wavelength, but the absolute value decays without oscillations. Inset: The in-plane spatial Fourier transform of the real and imaginary parts of the Green's dyadic. The imaginary part has a peak at the surface phonon polariton wavevector $\kappa_{spp}=(\omega_r/c)\sqrt{\varepsilon_2(\omega_r)/[\varepsilon_2(\omega_r)+1]}$ (vertical dashed line).}
 \label{fig:T12_sic}
 \end{center}
\end{figure}

To investigate how the surface phonon polaritons affect interparticle heat transfer close to a SiC surface, we plot in Fig. \ref{fig:T12_sic}(a) the transmission functions for two nanoparticles at a distance of $1$ $\mu$m from a SiC surface, as shown in the inset. The transmission functions are again evaluated at the energy $\hbar \omega_r=115.6$ meV. For $b \gtrsim 5$ $\mu$m, the interparticle transmission function $\ca{T}_{12}$ for particles close to a surface (solid line) is clearly enhanced compared to the free space value (dashed line) by the coupling to the surface phonon polaritons. Similarly to the case of nanoparticles in a cavity, the enhanced interparticle transmission is in most cases negligible compared to the strong coupling $\ca{T}_{2,\textrm{rad}}$ to the environment radiation which also increases due to the coupling to surface phonon polaritons. 

In contrast to Fig. \ref{fig:T12_b}, the interparticle transmission does not oscillate as a function of particle separation. This can be understood by noting that since the phonon polaritons carrying the heat are excited at a single wavevector $\kappa$, no interference oscillations can be expected. To show explicitly that the interparticle heat transfer enhancement is mainly due to propagating surface phonon polaritons, the $xx$-component of the scattering part of the interparticle Green's dyadic $\gem^{s}_{12}(\omega)$ is plotted in Fig. \ref{fig:T12_sic}(b). The $xy$- and $yy$-components of the Green's dyadics are similar, and the $xx$-component is chosen only for illustration. The phase of the Green's dyadic can be seen to rotate at the wavelength $\lambda \approx 7.5$ $\mu$m corresponding to the wavevector $\kappa\approx 0.84$ $\mu$m$^{-1}$, which agrees with the phonon polariton in-plane wavevector $\kappa_{\textrm{spp}}=(\omega/c) \sqrt{\varepsilon_2(\omega)/[\varepsilon_2(\omega)+1]}\approx (0.83+0.033i)$ $\mu$m$^{-1}$. The $90^{\circ}$ phase difference of the real and imaginary parts of Green's dyadic physically means that the phase of the electric field at the location of the particle $1$ rotates as a function of separation $b$ from dipole $2$. Since the polaritons propagate freely on the surface, no interference oscillations can be seen either in the absolute value of the Green's dyadic or in the transmission function. Interference effects could be observed by suitably structuring the surface such that phonon polaritons could be reflected by the obstructions or by replacing the surface by a thin film supporting two branches of surface phonon polaritons \cite{burke86}.  

\section{Summary and outlook}

We developed a microscopic dipole oscillator model to describe electromagnetic heat transfer between an arbitrary number of objects each described as a single oscillating dipole or more generally as a collection of dipoles in the spirit of the discrete dipole approximation. The dynamics of each dipole was described by the Langevin theory accounting for the dissipation and noise induced by the thermal fluctuations in the dipole moment. The formulation enables a physically insightful approach to model the energy transfer and results in heat transfer rates that are fully analogous with the Landauer-B\"uttiker formulas of electron and phonon transfer. We also expressed the energy transmission functions in terms of purely optical properties, i.e. the polarizabilities of the dipoles and the electromagnetic Green's dyadic. 

We applied the formalism to calculate heat transfer rates between silicon carbide nanoparticles placed in a microcavity, showing that the interparticle heat transfer rate can be enhanced by orders of magnitude as compared to a similar nanoparticle configuration in free space. The coupling of the background thermal field is, however, also enhanced by the microcavity and typically dominates the net energy transfer rate of a nanoparticle with its surroundings. As a function of in-plane separation of two nanoparticles in a microcavity, the interparticle heat current also oscillates due to the interference of the excited cavity modes. 

The quantum Langevin approach presented in this paper opens up new possibilities for modeling energy transfer and the coupling of different energy transfer mechanisms in nanostructures. For instance, coupling the presently studied system to phonon heat transfer model described by similar Langevin equations of motion will enable a detailed description of energy transfer by a coupled photon and phonon system. Further coupling of phonon and electron baths would capture inelastic energy transfer between the lattice vibrations and electrons, thereby providing a very thorough picture of energy transport and conversion processes in nanostructures, with numerous applications in the modeling of future energy technological devices.

\section{Acknowledgements}
We thank O. Heikkil\"a for useful discussions especially on the calculation of the cavity Green's dyadic. We acknowledge the Aalto Science-IT project and Finnish IT Center for Science for computer time. The work is in part funded by the Academy of Finland and the AEF research program of Aalto University and by the Graduate School in Electronics, Telecommunications and Automation (GETA).

\appendix
\section{Derivation of Eq. \eqref{eq:landauer_formula}}
\label{sec:appendix1}
In this appendix, we calculate the thermal average of the Langevin bath current $Q_i^{\textrm{bath}}$ defined in Eq. \eqref{eq:qi_def} using the solution \eqref{eq:usol} of the dipole displacements and the correlation functions \eqref{eq:xiixij} and \eqref{eq:ebgebg1}. For convenience, we define the reduced correlator $\langle \hat A\hat B \rangle_{\omega}$, valid for any functions $\hat{A}$, $\hat{B}$, through the relation
\begin{equation}
 \langle \hat A(\omega) \hat B(\omega') \rangle = 2\pi\delta(\omega+\omega') \langle \hat A\hat B\rangle_{\omega}.
\end{equation}
We separate the energy current $Q_i^{\textrm{bath}}$ into two terms as $Q_i^{\textrm{bath}}=-Q^{(1)}_i+Q^{(2)}_i$, where $Q^{(1)}_i=\dot{\bu}_i \cdot\xi_i $ and $Q^{(2)}_i=m\gamma\dot{\bu}_i$. The rate of work done by the stochastic force $\xi_i$ on the dipole at site $i$ is
\begin{alignat}{2}
 \langle Q^{(1)}_i \rangle &= \langle \dot{\bu}_i (t) \cdot\xi_i (t) \rangle \\
  &=  \left\langle\int \frac{d\omega}{2\pi} \frac{d\omega'}{2\pi}  (-i\omega)\hat\bu_i(\omega)^T \hat\xi_i(\omega') e^{-i\omega t-i\omega' t} \right\rangle \\
  &=  i \int \frac{d\omega}{2\pi}  \omega \sum_j \textrm{Tr} \left\langle \left[\hat \xi_j^T +q\Eenvhat(\br_j) ^T\right] \bG_{ji}^T  \hat\xi_i  \right\rangle_{\omega}, 
\end{alignat}
where we substituted the solution \eqref{eq:usol} for the dipole dynamics. We also introduced the trace, which allows us to cycle the position of the term inside braces:
\begin{alignat}{2}
  \langle Q^{(1)}_i \rangle &=  i \int \frac{d\omega}{2\pi}  \omega \sum_j \textrm{Tr} \left\{ \bG_{ji}\pom^T   \left\langle \hat \xi_i \left[\hat \xi_j^T+q\Eenvhat(\br_j) ^T\right\} \right\rangle_{\omega} \right]\\
  &=  i \int \frac{d\omega}{2\pi} \hbar \omega  \textrm{Tr} \left[\bG_{ii} \pom \Gamma^{\textrm{bath}}_i\pom \right] \fbi	.
\end{alignat}
Here we used Eqs. \eqref{eq:xiixij} and \eqref{eq:xiiebg}. Since the Green's function is a real function in time-domain, the real part of $\bG_{ii}(\omega)$ is an even function of frequency and since $\Gamma^{\textrm{bath}}_i\pom=2m\gamma_i \omega \unitdyadic$ and $\omega[f_B(\omega,T_i)+1/2]$ are both odd functions of frequency, only the imaginary part of $\bG_{ii}(\omega)$ survives the integration. The same result would follow directly by considering the symmetrized function $(\dot{\bu}_i\cdot\xi_i+\xi_i\cdot\dot{\bu}_i)/2$ instead of $\dot{\bu}_i\cdot\xi_i$, which would be necessary for Heisenberg operators \cite{saaskilahti13}. We get
\begin{alignat}{2}
  \langle Q^{(1)}_i \rangle&= -\int \frac{d\omega}{2\pi}\hbar  \omega  \textrm{Tr} \left\{\textrm{Im}[\bG_{ii}\pom ]\Gamma^{\textrm{bath}}_i\pom \right\} \fbi .
\end{alignat}
Now we apply the identity
\begin{equation}
 \textrm{Im}[\bG\pom ] = -\frac{1}{2} \bb{G}\pom\left[\Gamma^{\textrm{bath}}\pom+\Gamma^{\textrm{rad}}(\omega) \right]\bb{G}\pom^{\dagger}, \label{eq:gammaidentity}
\end{equation}
where the block-diagonal matrices $\Gamma^{\textrm{bath}}(\omega)= \textrm{diag}[\Gamma_1^{\textrm{bath}}\pom,\dots,\Gamma_N^{\textrm{bath}}\pom]$ and $\Gamma^{\textrm{rad}}(\omega)=2q^2 \omega^2 \mu_0 \textrm{Im}[ \gem\pom]$ were defined in Sec. \ref{sec:equationofmotion}. We get
\begin{alignat}{2}
  \langle Q^{(1)}_i \rangle 
  &=  \int_0^{\infty} \frac{d\omega}{2\pi} \hbar \omega \sum_{j=1}^N \textrm{Tr} \left[\bG_{ij}\pom \Gamma_j^{\textrm{bath}}  \pom \bG_{ji}(\omega)^{\dagger} \Gamma^{\textrm{bath}}_i\pom \right] \fbi \notag  \\
  &\quad + \int_0^{\infty} \frac{d\omega}{2\pi} \hbar \omega  \textrm{Tr} \left\{ \left[\bG\pom \Gamma^{\textrm{rad}} \pom\bG(\omega)^{\dagger} \right]_{ii} \Gamma^{\textrm{bath}}_i\pom \right\} \fbi.
\end{alignat}
Here we noted that the integrands are even functions in $\omega$.

The power absorbed by the bath due to the friction force is 
\begin{alignat}{2}
 \langle Q^{(2)}_i \rangle &= \langle m\gamma\dot{\bu}_i(t)^2 \rangle \\
  &=  \int \frac{d\omega}{2\pi} m \gamma \omega^2 \textrm{Tr} \langle \hat \bu_i^T \hat \bu_i \rangle_{\omega} \\
  &= \int \frac{d\omega}{2\pi} m\gamma\omega^2  \sum_{j,k=1}^N \textrm{Tr} \left\langle \left[\hat\xi_j^T+q\Eenvhat(\br_j)^T\right] \bG_{ji}^T \bG_{ik} \left[\hat\xi_k+q\Eenvhat(\br_k)\right] \right\rangle_{\omega} \\
  &=  \int \frac{d\omega}{2\pi}  m\gamma\omega^2  \sum_{j,k=1}^N \textrm{Tr}\left\{  \bG_{ji}\pom^T \bG_{ik}\pom^* \left\langle \left[\hat\xi_k+q\Eenvhat(\br_k)
  \right] \left[\hat\xi_j^T+q\Eenvhat(\br_j)^T\right] \right\rangle_{\omega} \right\} \\
  &= \int \frac{d\omega}{2\pi}  m\gamma \hbar \omega^2  \sum_{j=1}^N \textrm{Tr}\left[\bG_{ji}\pom^T \bG_{ij}\pom^* \Gamma_j^{\textrm{bath}}\pom \right] \fbj  \notag \\
  &\quad + \int \frac{d\omega}{2\pi}  m\gamma \hbar \omega^2 \sum_{j,k=1}^N \textrm{Tr}\left[\bG_{ji}\pom^T \bG_{ik}\pom^* \Gamma_{kj}^{\textrm{rad}}\pom \right] \fbbg. \\
    &= \frac{1}{2} \int \frac{d\omega}{2\pi} \hbar \omega  \sum_{j=1}^N \textrm{Tr}\left[\Gamma_i^{\textrm{bath}}\pom \bG_{ij}\pom^* \Gamma_j^{\textrm{bath}}\pom \bG_{ji}\pom^T\right] \fbj  \notag \\
    &\quad +\frac{1}{2} \int \frac{d\omega}{2\pi} \hbar \omega \sum_{j,k=1}^N \textrm{Tr}\left[\Gamma_i^{\textrm{bath}}\pom \bG_{ik}\pom^* \Gamma_{kj}^{\textrm{rad}}\pom \bG_{ji}\pom^T \right] \fbbg. 
\end{alignat}
Taking the complex conjugate gives finally
\begin{alignat}{2}
 \langle Q^{(2)}_i \rangle 
    &= \int_0^{\infty} \frac{d\omega}{2\pi} \hbar \omega  \sum_{j=1}^N \textrm{Tr}\left[\Gamma_i^{\textrm{bath}}\pom \bG_{ij}\pom \Gamma_j^{\textrm{bath}}\pom \bG_{ji}\pom^{\dagger}\right] \fbj  \notag \\
    &\quad + \int_0^{\infty} \frac{d\omega}{2\pi} \hbar \omega \textrm{Tr}\left\{\Gamma_i^{\textrm{bath}}\pom [\bG\pom \Gamma^{\textrm{rad}}\pom \bG\pom^{\dagger}]_{ii} \right\} \fbbg. 
\end{alignat}
The second term is absent, if the environment field is absent. The total heat current flowing to the bath is $\langle Q^{\textrm{bath}}_i \rangle = \langle Q^{(2)}_i \rangle - \langle Q^{(1)}_i \rangle$, so we get Eq. \eqref{eq:landauer_formula}.


\end{document}